\documentclass[10pt]{iopart}

\usepackage{microtype, textcomp}
\usepackage[abbreviations = false, detect-all, list-units = single]{siunitx}
\usepackage{amssymb,bm}

\usepackage{cite}
\usepackage{subfig}
\usepackage[inline]{enumitem}
\usepackage{url}
\urldef{\ourhomepage}\url{http://www.lens.unifi.it/quantum-nanophotonics/mcplusplus/lut}
\urldef{\oursourcecode}\url{http://www.lens.unifi.it/quantum-nanophotonics/mcplusplus}

\usepackage{graphicx}

\begin{document}
\title{Deducing effective light transport parameters in optically thin systems}
\author{G Mazzamuto$^{1,2}$, L Pattelli$^{1}$, C Toninelli$^{1,2,3}$, D S Wiersma$^{1,4}$}

\address{$ˆ1$ European Laboratory for Non-linear Spectroscopy (LENS), Universit\`{a} di Firenze, 50019 Sesto Fiorentino (FI), Italy}
\address{$ˆ2$ CNR-INO, Istituto Nazionale di Ottica, Via N. Carrara 1, 50019 Sesto Fiorentino (FI), Italy}
\address{$ˆ3$ QSTAR, Largo Enrico Fermi 2, 50125 Firenze, Italy}
\address{$^4$ Department of Physics, Universit\`{a} di Firenze, 50019 Sesto Fiorentino (FI), Italy}
\ead{pattelli@lens.unifi.it}

\begin{abstract}
We present an extensive Monte Carlo study on light transport in optically thin slabs, addressing both axial and transverse propagation. We completely characterize the so-called ballistic-to-diffusive transition, notably in terms of the spatial variance of the transmitted/reflected profile. We test the validity of the prediction cast by diffusion theory, that the spatial variance should grow independently of absorption and, to a first approximation, of the sample thickness and refractive index contrast.
Based on a large set of simulated data, we build a freely available look-up table routine allowing reliable and precise determination of the microscopic transport parameters starting from robust observables which are independent of absolute intensity measurements.  We also present the Monte Carlo software package that was developed for the purpose of this study.
\end{abstract}

\noindent{\it diffusion, turbid media, photon migration, light propagation in tissues, Monte Carlo simulations, parallel processing, look-up table\/}
\maketitle

\section{Introduction}
Light represents a powerful and versatile tool to investigate optical properties of complex media such as biological tissues\cite{tuchin2007tissue}. The radiative transport features of a disordered sample are indeed deeply related to its microscopic structural\cite{friebel1999optical, backman1999polarized, burresi2014bright} and chemical properties\cite{mourant1998mechanisms, zonios2001skin}, which makes light propagation studies a desirable non-invasive and non-destructive diagnostic tool in the most diverse applications.

In this context, the Diffusion Approximation (DA) represents an incredibly robust theoretical framework. As such it provides a complete set of straightforward analytical expressions to describe light transport both in the spatial and temporal domain, thus allowing to solve the inverse transport problem and eventually retrieve the microscopic parameters. Nonetheless it significantly fails in those circumstances where its fundamental assumption of almost isotropic radiance is not verified, a condition that is truly fulfilled only asymptotically in space and time. As a consequence, a vast effort is devoted to characterize its precision, accuracy and validity range, which are known to become defective in low albedo scattering systems and, most importantly, media whose extension is not large enough to allow the onset of a multiple scattering regime.
Among these, thin slabs and membranes represent a very common and relevant geometry, especially in the biological field\cite{cheong1990review}, where many tissues such as the ocular \textit{fundus}\cite{hammer1995optical}, vascular walls\cite{chan1996effects}, cellular cultures\cite{backman1999polarized}, skin dermis\cite{du2001optical}, skull bones \cite{ugryumova2004measurement} or dental enamel\cite{darling2006light} --- to name a few --- are inherently available only within narrow ranges of minute sizes and thicknesses. For these systems, experimental data evaluation must rely on more accurate methods such as refined approximations of the Radiative Transfer Equation\cite{elaloufi2002time, xu2002photon, liemert2010analytical, liemert2013exact} or Monte Carlo simulations\cite{wang1995mcml, alerstam2008parallel}, the latter providing an exact solution to the Radiative Transfer Equation for any sample geometry, given that large enough statistics are collected.

Nonetheless, due to its simplicity, the diffusion approximation still retains a large appeal, and great efforts are constantly made to extend its validity range introducing all sorts of minor modifications\cite{lemieux1998diffusing, venugopalan1998radiative, garofalakis2004characterization}. Even at its standard formulation, diffusion theory casts an incredibly simple prediction on \emph{transverse} transport which could be profitably applied in a thin slab geometry, since boundary and confinement effects are less relevant along the slab's main extension. A light beam impinging on a disordered slab is usually transmitted with an enlarged spatial profile which, according to the diffusion approximation, is gaussian with a standard deviation growing linearly in time as $w^2 (t) = 4Dt$ with a slope determined by the diffusion coefficient. This quantity is more generally referred to as the Mean Square Width (MSW), and can be defined for an arbitrary intensity distribution $I(\rho, t)$ through the relation
\begin{equation}
w^2(t) = \frac{\int \rho^2 I(\rho, t) \, \text{d}\rho}{\int I(\rho, t) \, \text{d}\rho}.
\label{eq:MSW}
\end{equation}
Experimentally, the temporal evolution of the mean square width can be measured from a collection of discrete, spatio-temporally resolved profiles $I(\rho, t_i)$\cite{sperling2012direct, pattelli2015visualizing}. Its linear increase can be considered as a signature for transverse diffusion, with the obtained slope representing a valuable observable since the MSW is remarkably unaffected by the presence of absorption (which cancels out exactly at any $t_i$). Even more surprisingly, the diffusion approximation predicts the mean square width slope to be independent from both the slab thickness and its refractive index contrast with the surrounding environment. These properties are unique to the MSW, whereas typically any other observables depend critically on absorption, slab thickness and boundary conditions\cite{zhu1991internal, bouchard2010reference}. 
Nevertheless, transverse transport has been so far largely disregarded, supposedly because it was hardly accessible experimentally.

The aim of this study is to characterize light transport in the infinite slab geometry over a wide range of optical parameters, testing for the first time to our knowledge how the linear prediction from the diffusion approximation performs as the optical thickness of the sample decreases. In this respect, our results fill the gap with the already extensive characterization available in the literature regarding the breakdown of the diffusion model along the propagation axis\cite{mackintosh1989diffusing, yoo1990does, yoo1990time, alerstam2008improved, bouchard2010reference, elaloufi2004diffusive, spinelli2007calibration, martelli2007calibration, svensson2013exploiting, pattelli2015visualizing}, finally providing the complete picture of the transition between the ballistic and diffusive regime. 
Moreover we tackle the problem of transverse transport in a confined geometry, which is one of fundamental interest especially when considering that this configuration allows the onset of multiple scattering also in semi-transparent media which are not usually associated with this transport regime\cite{pattelli2015diffusion, leonetti2011measurement}. Furthermore we envision that also from an experimental point of view our study will allow to exploit the large potential of novel optical investigation techniques based on the mean square width. To this purpose, we present a freely available look-up table (LUT) interface as a straightforward tool to fully take advantage of our simulated dataset and retrieve effective light transport parameters for optically thin systems, starting from robust observables which are independent of absolute intensity measurements.

\section{Results and Discussion}
In the first part of our work we perform a systematic Monte Carlo study over a range of different optical properties aimed at testing the validity of the diffusion approximation (DA). In particular we test the linear prediction $w^2 (t) = 4Dt$, with $D = D_\text{DA} = l_\text{t} v/3$, against $w^2 (t)$ values obtained using \eref{eq:MSW} on a single photon basis. Assuming a slab thickness of $L=\SI{1}{\milli\meter}$ we simulated different optical thicknesses (OT) ranging from \numrange{1}{10} by varying the transport mean free path $l_\text{t}$ between \SIlist{0.1;1}{\milli\meter}. Each OT simulation has been run for 11 different values of the anisotropy factor $g$, defined as the average cosine of the scattering angle $\theta$, between \numlist{0;0.99} and 16 values of the refractive index contrast $n$ ranging from \numrange{0.6}{2.2}, for a grand total of 2816 simulations of \num{e9} photons each (see Methods). Photons are injected normally to the slab plane from a $\delta(r)\delta(t)$ pencil beam source and collected at the opposite surface (i.e.~we only consider transmitted intensity). For the sake of convenience, since Fresnel reflection coefficients depend solely on the relative refractive index contrast $n = n_\text{in} / n_\text{out}$, we kept $n_\text{in} = 1$ constant while varying $n_\text{out}$ in order to have a consistent time scale over the whole set of simulations. The real time scale of any single simulation can be recovered by simple multiplication by the actual value of $n_\text{in}$.

\subsection{Mean square width expansion}
\begin{figure}[htbp]
\centering
\subfloat[][]{
\includegraphics{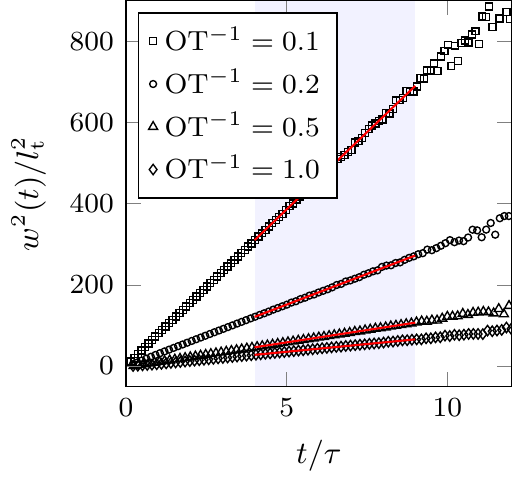}
}\hspace*{\fill}
\subfloat[][]{
\includegraphics{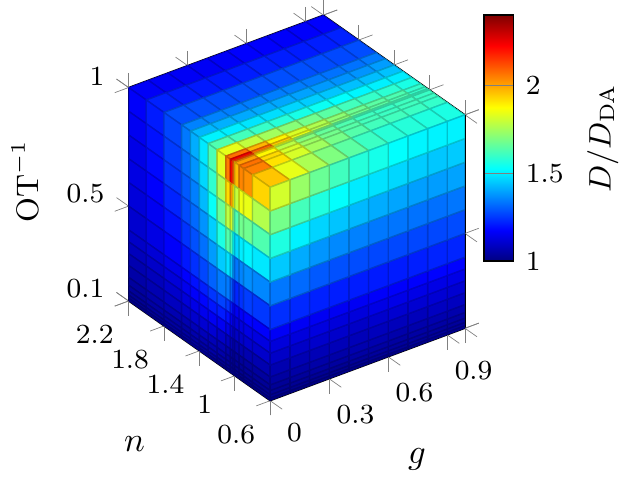}
}\\
\subfloat[][]{
\includegraphics{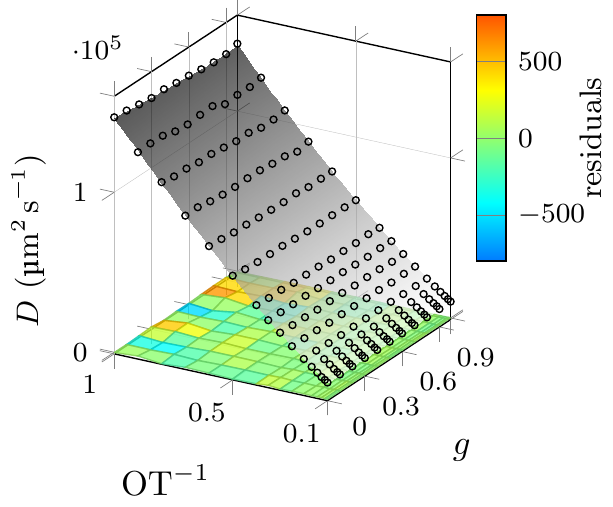}
}\hspace*{\fill}
\subfloat[][]{
\includegraphics{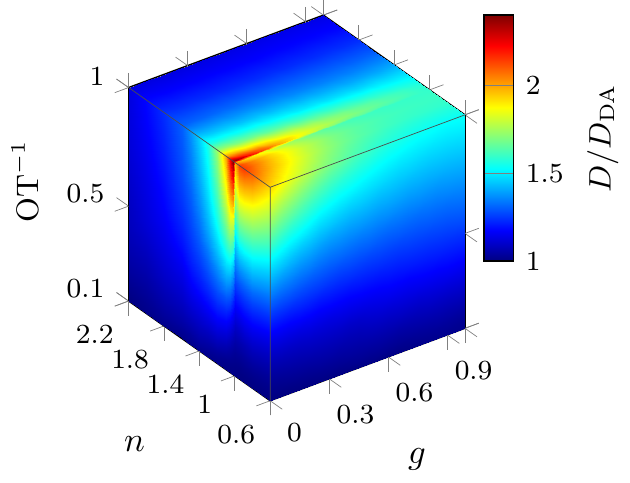}
\label{fig:interpolatedDCube}
}\\
\caption{Procedure followed to generate the hyper-surface of relative MSW slope deviations. (a) Subset of simulated time-resolved MSW for $n=1.4$, $g=0.9$ and different values of $\text{OT}^{-1} = l_\text{t}/L_0$. The MSW slope is fitted for each curve by applying a linear fitting model over a temporal window ranging from 4 to 9 decaying lifetimes, which in turn have been estimated by time-resolved curves (not shown). (b) Final hyper-surface showing the $D/D_\text{DA}$ ratio over the whole range of simulated parameters. (c) Each simulated $n$-slice ($n=1.4$ shown) is processed through a Loess fitting routine (range parameter set to \num{0.25}). (d) Smoothed slices are eventually put together in order to carry a gridded interpolation along the refractive index contrast axis.}
\label{fig:Dcube}
\end{figure}

\Fref{fig:Dcube}a shows a subset of simulated mean square width data obtained for typical optical properties of relevance in the bio-optical field ($n=1.4$, $g=0.9$), exhibiting a perfectly linear increase also in the optically thinnest case. As previously specified, mean square width values are exactly independent of absorption, which thus has been excluded from the simulations. The retrieved values of $D$ evaluated as 1/4 of the variance slope have been divided by the expected value $D_\text{DA} = l_\text{t} c/3$ and arranged in a hyper-surface of relative deviations (\fref{fig:Dcube}b). The obtained volume is sampled in a discrete set of points in the ($n$, $g$, $\text{OT}^{-1}$) space where the single simulations have been performed. Noise coming either from limited statistics or fitting uncertainties is largely reduced by applying a local regression algorithm using weighted linear least squares and a 2\textsuperscript{nd} degree polynomial model as provided by the Loess MATLAB model (\fref{fig:Dcube}c).
This allows to eventually obtain an accurate, smoothly sampled volume suitable for finer interpolation (\fref{fig:Dcube}d).

A few comments are in order.
Firstly, the present investigation is intended to focus on long-time/asymptotic transport. To this purpose, the diffusion coefficient $D$ has been evaluated by the linear slope of the mean square width (MSW) in a time window ranging from 4 to 9 decaying lifetimes, as determined from time-resolved curves. Depending mainly on the optical thickness of the sample, there is an early-time range where the MSW exhibits a super-linear increase. We carefully checked that the aforementioned fitting time range was always largely excluding such non-linear time range in order to address safely the asymptotic slope, as confirmed for example in \fref{fig:Dcube}a.

Secondly, it is well known that most biological soft tissues share a refractive index equal or close to $n_\text{in}=1.4$\cite{bolin1989refractive}. This is supposedly the reason why refractive index variations have so far been disregarded in similar multi-parameter investigations\cite{pifferi1998real, dam2000multiple, rajaram2008lookup, hennessy2013monte}. Nonetheless, we included the refractive index contrast as a simulation parameter because, especially in the case of thin slabs, the range of interest for $n$ is undoubtedly wider, spanning from well below 1 to as high as 2. The first case for small $n$ is of interest for cases where specimens are enclosed in glass slides, or laid or immersed in different substrates/solutions, whereas the high values for $n$ have been included envisioning possible applications of our study to metal oxides and similar highly scattering materials, which are extremely relevant, for instance, for coatings and in photovoltaics.

Looking at the obtained data, two features are immediately noticeable. Firstly, the diffusion approximation appears to always underestimate the actual spreading rate, of course recovering agreement for higher optical thicknesses as expected. A second, finer feature occurs in the close proximity of $n=1$, particularly evident at low $g$ and OT values. Both these features arise from the interplay between geometric and boundary conditions. In particular, we found that the presence of internal reflections in a thin layer geometry helps to selectively hold inside the slab those photons that happen to draw statistically longer steps, as we discuss in more detail in a related work\cite{pattelli2015diffusion}. To the purpose of this study, it is worth noting that the mean square width slope exhibits a distinct pattern of characteristic deviations from the diffusion approximation, which can therefore be exploited as a guide to unambiguously retrieve the \emph{intrinsic} microscopic transport properties of a given sample.

\subsection{Decay lifetime}
It is interesting to consider also how the decaying lifetime is affected by different configurations. In the diffusion approximation, for a non-absorbing medium we have:
\begin{equation}
\tau_\text{DA} = \frac{L_\text{eff}^2}{\pi^2 D},
\label{eq:tauDA}
\end{equation}
where $L_\text{eff} = L + 2z_\text{e}$ is the effective thickness of the medium and $z_\text{e}$ represents the extrapolated boundary conditions for a given refractive index contrast.

\begin{figure}[htbp]
\centering
\subfloat[][]{
\includegraphics{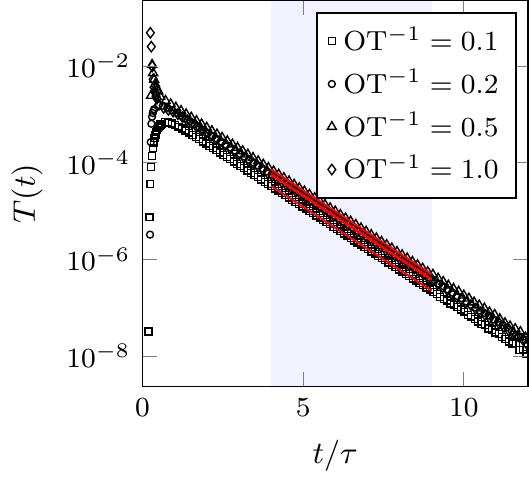}
}\hspace*{\fill}
\subfloat[][]{
\includegraphics{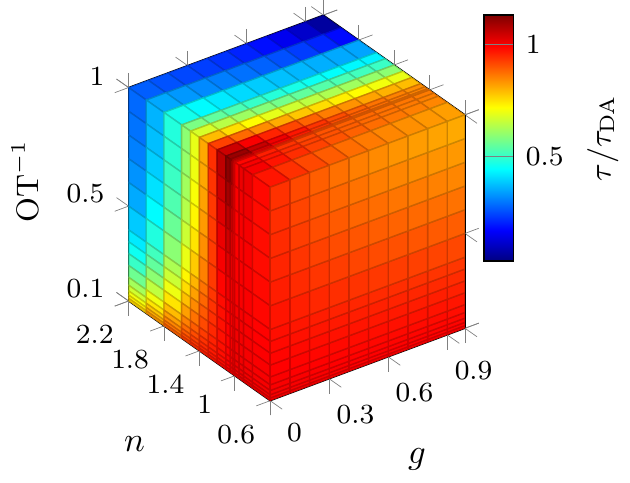}
}\\
\subfloat[][]{
\includegraphics{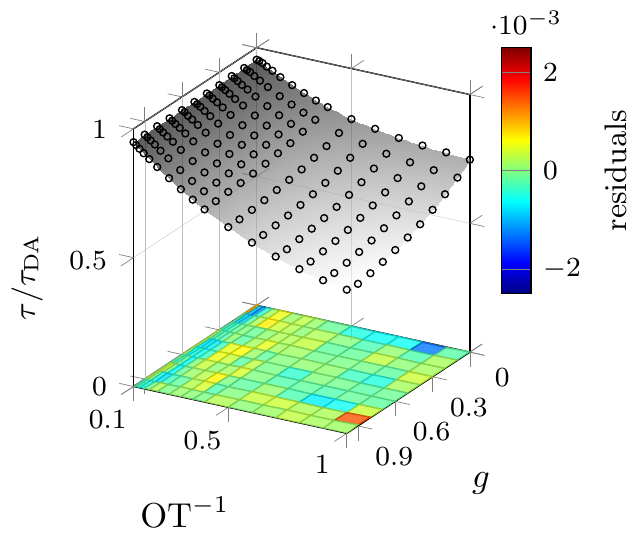}
}\hspace*{\fill}
\subfloat[][]{
\includegraphics{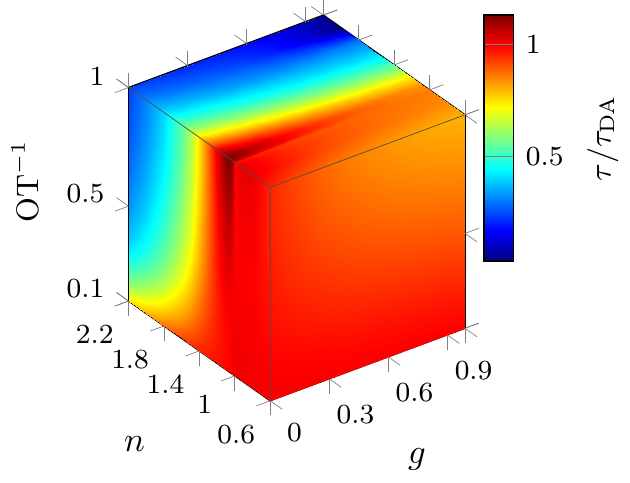}
\label{fig:interpolatedLifetimeCube}
}
\caption{Procedure followed to generate the hyper-surface of relative lifetime deviations. (a) Subset of simulated time-resolved transmittance for $n=1.4$, $g=0.9$ and different values of $l_\text{t}/L_0$. The lifetime is fitted for each curve by applying an exponential fitting model over a temporal window ranging from 4 to 9 decaying lifetimes, a parameter which is iteratively estimated in a multi-step process. (b) Final hyper-surface showing the $\tau/\tau_\text{DA}$ ratio over the whole range of simulated parameters. (c) Each simulated $n$-slice ($n=1.4$ shown) is processed through a Loess fitting routine (range parameter set to \num{0.25}). (d) Smoothed slices are eventually put together in order to carry a gridded interpolation along the refractive index contrast axis.}
\label{fig:lifetimecube}
\end{figure}

Figures \ref{fig:lifetimecube}a and b respectively show a typical set of time-resolved transmittance decays and the hyper-surface of relative deviations from the diffusion approximation prediction.
Two main features are worth commenting when comparing to the previous results for the mean square width slope. First of all the obtained lifetime deviations are more significant, reaching down to \SI{20}{\percent} of the expected value for the highest values of $g$ and $n$. It is indeed known that $n>1$ refractive index contrasts are harder to be taken into account even when applying appropriate boundary corrections and even at higher optical thicknesses. Secondly, deviations in both directions are possible, with the $\tau/\tau_\text{DA}$ ratio taking values both above and below \num{1}. This might help explaining some experimental evidences obtained in thin disordered samples that are still debated at a fundamental level\cite{kop1997observation, rivas2001static, ramakrishna1999diffusion, gopal2001photon, zhang2002wave, elaloufi2002time}, as further discussed in a related paper\cite{pattelli2015diffusion}.
Our findings stress the importance of an accurate and precise modeling of the index contrast, which we think has been often overlooked, for example when a symmetric averaged contrast is used to model asymmetric experimental configurations\cite{ramakrishna1999diffusion, zhang2002wave, elaloufi2002time}.

Despite the vast literature regarding the validity range of the diffusion approximation in the time domain\cite{mackintosh1989diffusing, yoo1990does, yoo1990time, alerstam2008improved, bouchard2010reference, elaloufi2004diffusive, spinelli2007calibration, svensson2013exploiting}, a comprehensive understanding of the interplay between optical thickness, refractive index contrast and absorption is still a debated topic.
It is a common assumption that the diffusion approximation fails gradually with decreasing optical thickness, with $\text{OT} = 8$ being customarily considered as the lower threshold under which the introduced error starts to be significant\cite{elaloufi2004diffusive}. Nevertheless, it was recently shown how a non-absorbing, $\text{OT} > 8$ slab sample with $n \sim 1.5$ exhibits a transmittance lifetime such that the diffusion approximation is unable to provide \emph{any} real solution at all\cite{pattelli2015visualizing}, thus suggesting that the breakdown of the diffusion approximation might step in abruptly depending on the interplay between different parameters other than the optical thickness.
In the same work, by resorting to Monte Carlo simulations it was possible to correctly retrieve the transport mean free path and absorption coefficient beyond the evaluation capabilities of the diffusion approximation. Notably, the observed deviations from both the experimental decay lifetime and the mean square width expansion are in perfect agreement with our simulations.

\subsection{Absorption}
By combining our mean square width and decay lifetime analysis some unique insight on the effects of absorption is gained. As already mentioned, absorption has not been considered as a simulation parameter since its effects can be accounted for \emph{exactly} both in $w^2 (t)$ and $\tau$. Concerning the mean square width, its value is completely unaffected at any time by the presence of any amount of absorption, while the asymptotic decay lifetime is expected to shift exactly to
\begin{equation}
\frac{1}{\tau} \quad \rightarrow \quad \frac{1}{\tau} + \mu_\text{a}v ,
\label{eq:tau_absorption}
\end{equation}
$\tau$ being the lifetime in the non-absorbing case.
This is equivalent to saying that whenever we know the specific intensity $I(\bm{r}, t, \mathbf{s}; \mu_\text{a} = 0)$ at a given position $\bm{r}$, time $t$ and direction $\mathbf{s}$, then
\begin{equation}
I(\bm{r}, t, \mathbf{s}; \mu_\text{a}) = \mathrm{e}^{-\mu_\text{a} v t}\, I(\bm{r}, t, \mathbf{s}; 0)
\label{eq:LBlaw}
\end{equation}
is the solution when the absorption coefficient (independent of $\bm{r}$) is $\mu_\text{a}$.

The problem with absorption is that both scattering and absorption can deplete specific intensity from a given position, time and direction (an effect sometimes referred to as absorption-to-scattering cross-talk). Hence, retrieving its unknown value from experimental data has been to date a challenging task.
Besides that, it is often reported that the diffusion approximation is expected to hold only for weakly absorbing media since the onset of the properly diffusive regime requires long trajectories to contribute dominantly to transport properties, whereas these are selectively suppressed by absorption\cite{alerstam2011optical}. This explains why absorption is often considered as a major hindrance in the correct assessment of transport properties\cite{fishkin1996gigahertz, martelli2000accuracy, ntziachristos2001accuracy}, if not even an invalidating condition for certain optical parameter measurements\cite{wiersma1997localization, scheffold1999localization, wiersma1999reply}.
For this reason, techniques capable of directly accessing the mean square width recently aroused a great deal of interest given the absorption-independent nature of the variance expansion\cite{hu2008localization, cherroret2010transverse, sperling2012direct}. From our perspective, the full potential of MSW measuring techniques is still to be fully unraveled, and will eventually play a key role among the most accurate characterization techniques of both scattering \emph{and} absorption properties, as we will demonstrate in the following section.

\subsection{Look-up table approach}
In the past years, increasing availability of massively parallel computation capabilities is fostering Monte Carlo simulations as a viable method to solve the inverse transport problem, i.e.~retrieving the microscopical properties that give rise to certain temporal or spatial profiles. Two different Monte Carlo-based approaches are represented by fitting and look-up table (LUT) routines respectively. The former would in principle require to solve iteratively the forward problem until some convergence condition between experimental and simulated data is satisfied. Due to the impractical computational load, fitting routines to date have exploited rescaling properties of the radiative transfer equation to adapt a limited set of pre-simulated Monte Carlo data to experimental measurements\cite{graaff1993condensed, hammer1995optical, kienle1996determination, pifferi1998real, xu2006investigation, alerstam2008improved, alerstam2008white, martinelli2011analysis}. In order to limit the occurrence of ``scaling artifacts'', rescaling must be typically performed on a single photon basis, thus requiring to store each exit time and position separately\cite{xu2006investigation, alerstam2008white}. Bin-positioning strategies are also known to represent a possible source of artifacts, requiring complex correction strategies to be deployed\cite{martinelli2011analysis}. Finally, while for the semi-infinite geometry a single dimensionless Monte Carlo simulation can be rescaled both in terms of absorption (exploiting \eref{eq:LBlaw}) and scattering mean free path, the computational cost increases in the case of finite thickness geometries, where different scattering mean free paths values must be simulated separately. This is why only few examples can be found in the literature dealing with this configuration\cite{pifferi1998real}.

Taking advantage of the large set of simulations performed, we developed a look-up table (LUT) routine to demonstrate the capabilities of a combined mean square width/lifetime approach. In the literature, several LUT approaches have been proposed, based on both experimental\cite{rajaram2008lookup} and simulated data\cite{nilsson1995measurements, roggan1995experimental, dam2000multiple, thueler2003vivo, dam2005real, palmer2006monte, karlsson2012inverse, hennessy2013monte}.
In contrast with a standard fit, look-up table routines rely on single scalar parameters directly linked to transport properties, such as, typically, the total amount of transmitted/ballistic/reflected light from a slab. This triplet of observables, often referred to as $T_\text{tot}$, $T_\text{coll}$ and $R_\text{tot}$, has been extensively exploited to retrieve optical parameters through Monte Carlo-LUT routines\cite{nilsson1995measurements, roggan1995experimental, dam2000multiple, thueler2003vivo, dam2005real, palmer2006monte, karlsson2012inverse, hennessy2013monte}, despite the fact that such absolute intensity measurements are extremely challenging\cite{dam2000multiple, dam2005real} and prone to unpredictable systematic errors\cite{karlsson2012inverse}. Moreover, while the natural propensity of Monte Carlo-based data evaluation techniques is towards the study of optically thin media, integrated transmittance/reflectance quantities do actually loose their usefulness as the sample thickness decreases\cite{hammer1995optical}, since the acquired signal will be eventually dominated by light that has been either specularly reflected or ballistically transmitted through the sample, thus carrying almost no information about its properties. Even more fundamentally, one of the main assumptions in the integrating sphere theory is that of a lambertian diffuse profile\cite{nilsson1995measurements}, which is clearly not holding for thinner systems. 

In this respect, for the first time to our knowledge, we designed a look-up table relying on observable quantities that are both free from any absolute intensity assessment and well into the multiple scattering regime, such as the asymptotic decay lifetime and the mean square width expansion slope. This offers several advantages over existing solutions:
\begin{enumerate}[label=\emph{\roman*}), leftmargin=*, nosep]
\item both the asymptotic decay lifetime and the mean square width slope can be measured without any reference to the excitation intensity, therefore there is no need to calibrate the source or the detector; lifetime determination is also not strictly connected to any particular detection geometry, which stands in contrast with other typical techniques often requiring a particular configuration of collection fibers, integrating spheres or angular measurements.
\item because of the asymptotic nature of both $\tau$ and $w^2 (t)$, the actual temporal response function or spatial excitation spot size are eventually irrelevant to their accurate determination.
\item precise determination of the origin of the time axis (i.e.~the exact time of pulse injection), while being dramatically relevant in many analogue situations, is here made completely irrelevant since both the decay lifetime and the linear increase of the mean square width do not exhibit any critical dependence on the exact delay at which they are determined, provided that it is sufficiently large.
\item with respect to Monte Carlo-based fitting routines, a look-up table routine is more suitable for real-time solving of the inverse problem since it does not involve any iterative procedure. While this guarantees better performances, on the downside we must note that it is less clear how to define the uncertainty of retrieved values. Typical approaches involve mapping the relative error of retrieved parameters over a broad range of independent simulations, in order to give a numerical estimate.
\item several issues typically associated to fitting routines are also obviated. Once that the two scalar parameters are calculated with the proper, original binning, they can be virtually rescaled arbitrarily without the risk of introducing any binning-related artifact.
\item the problem of correct bin positioning is also removed. Midpoint positioning adopted in our case represents the exact solution for the linear increase of the mean square width. While this is not the case for a monoexponential decay, it can be again shown trivially that midpoint positioning does leave the decay constant exactly unmodified.
\item as we will demonstrate in the following, a possible implementation of our LUT routine allows to retrieve $\mu_\text{a}$ and $l_\text{t}$. It must be nonetheless stressed again that none of the simulations that make up our look-up table include absorption, which stands in contrast with the customary practice of most LUT approaches demonstrated to date. This allows to deal with a simulation phase space of reduced dimensionality, which represents a huge saving on the computational burden of Monte Carlo simulations.
\item finally, since there is no need to directly simulate absorption \emph{nor add it after the simulation}, it is also not necessary to store exit times and positions on a single photon basis. This allows to hugely reduce the output size for each simulation and streamline its handling, thus allowing for better statistics to be collected.
\end{enumerate}

\begin{figure}[htbp]
\subfloat[][]{
\includegraphics{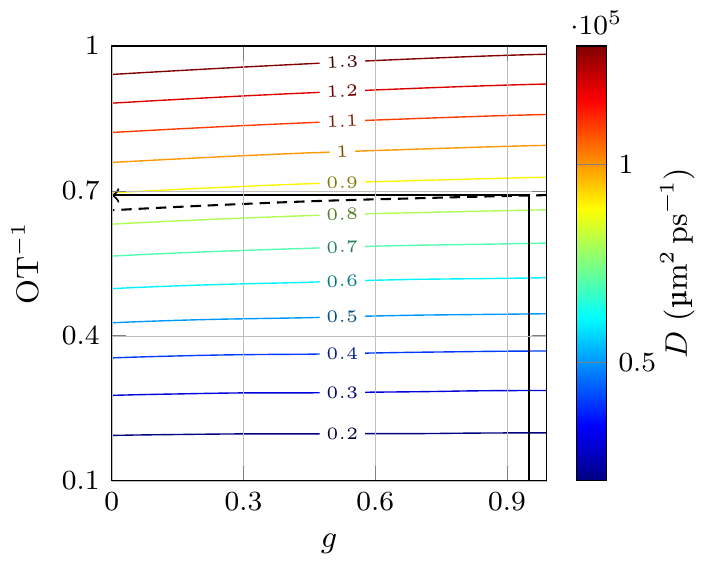}
}
\hspace*{\fill}
\subfloat[][]{
\includegraphics{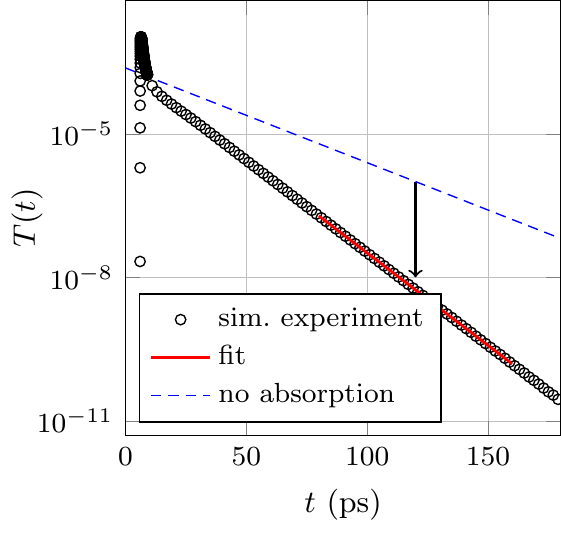}
}
\\
\subfloat[][]{
\includegraphics{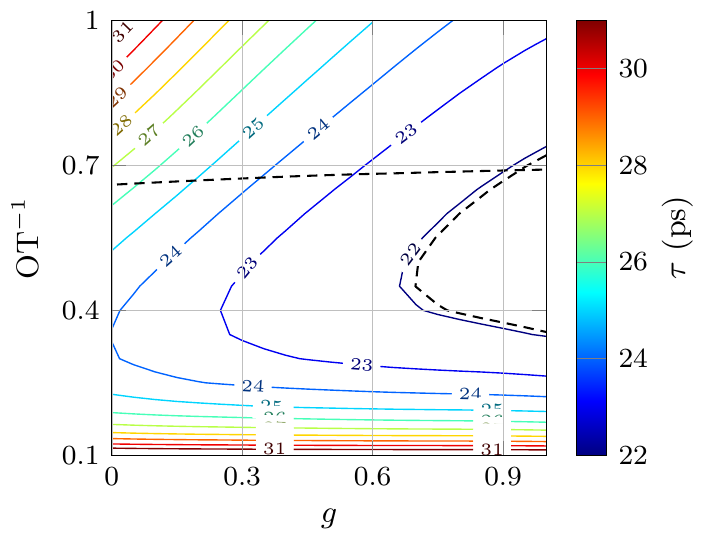}
\label{fig:isoTau}
}
\hspace*{\fill}
\subfloat[][] {
\includegraphics{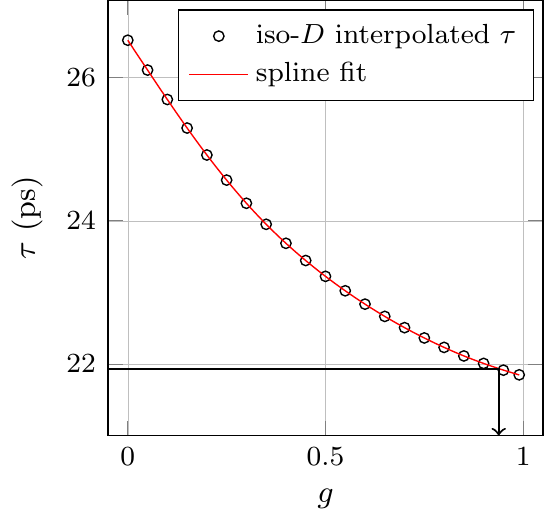}
}
\caption{Demonstration of the MC-LUT routine. We considered a sample with $L=\SI{1.3}{\milli\meter}$, $n=\num{1.38}$, $g=\num{0.95}$, $l_\text{s} = \SI{45}{\micro\meter}$ and $\mu_\text{a}$ respectively equal to \SI{0.2}{\per\milli\meter} for the $l_\text{t}$, $\mu_\text{a}$ retrieval demonstration (panels a and b) and $\mu_\text{a} = \SI{0}{\per\milli\meter}$ for the $l_\text{t}$, $g$ routine (panels c and d). Dashed line in the first contour plot marks the iso-$D$ curve calculated as $1/4$ of the ``experimental'' $w^2(t)$ slope. The same curve is traced also in third panel along with the corresponding iso-$\tau$ curve, with their intersection returning the estimated $g$ value (panel d).}
\label{fig:MCLUT}
\end{figure}

For the sake of simplicity, our present Monte Carlo-LUT demonstration is limited to the retrieval of pairs of transport parameters, e.g.~$l_\text{t}$ and $g$ (and therefore $l_\text{s}$) assuming that absorption is known, or $l_\text{t}$ and $\mu_\text{a}$ with known $g$, which is a realistic and common assumption in similar works, especially those involving biological samples\cite{kienle1996determination, pifferi1998real, dam2000multiple}. The effective refractive index and the thickness of the sample are also expected as input parameters.
To illustrate the steps involved in the llok-up table routine (\fref{fig:MCLUT}), we test the retrieval procedure on two simulated samples with $L=\SI{1.3}{\milli\meter}$, $n=\num{1.38}$, $g=\num{0.95}$, $l_\text{s} = \SI{45}{\micro\meter}$ and $\mu_\text{a}$ respectively equal to \SI{0.2}{\per\milli\meter} and \SI{0}{\per\milli\meter}. From these simulations we measure the mean square width and the decay lifetime, which we feed as test inputs into the LUT routine. First of all, the mean square width and lifetime hyper-surfaces must be rescaled both in space and time to match the target thickness and refractive index. The original simulations have been performed for a sample of thickness $L_0 = \SI{1}{\milli\meter}$ and unitary internal refractive index; dimensional analysis shows that eventually the mean square width and lifetime hyper-surfaces are to be rescaled by $L/(L_0 n_\text{in})$ and $n_\text{in} L / L_0$ respectively.

Let us consider first the general case of a medium with unknown scattering mean free path and absorption coefficient, but known $g$. Given the sample thickness and the internal/external refractive indices, it is possible to rescale the continuously interpolated version of the mean square width hyper-surface (\fref{fig:Dcube}d) in order to slice it at the given refractive index contrast $n$. The obtained 2D surface will feature an iso-level curve corresponding to the measured mean square width slope, which will basically give the expected $\text{OT}^{-1}$, i.e.~$l_\text{t}$, in a completely absorption-independent way. For the test sample described above, fitting the simulated mean square width values yields a slope of \SI{337750}{\square\micro\metre\per\pico\second}.
By intersecting this iso-level curve with the simulated value of $g=0.95$, we eventually retrieve the best $\text{OT}^{-1}$ estimate (\fref{fig:MCLUT}a). Nonetheless, even if the scattering anisotropy is not known \textit{a priori}, plugging reasonably bounded values into the routine helps getting an estimate of how an uncertainty on $g$ spreads over $l_\text{t}$ and eventually $\mu_\text{a}$.
Once that also $\text{OT}^{-1}$ is determined, it is sufficient to read the expected absorption-free lifetime value stored in $\tau(n, \text{OT}^{-1}, g)$ (shown in \fref{fig:MCLUT}b as a dashed blue line) from the interpolated lifetime hyper-surface (\fref{fig:interpolatedLifetimeCube}) and compare it directly to the measured value: the discrepancy between their reciprocal values will directly give $\mu_\text{a} c/n_\text{in}$ through \eref{eq:tau_absorption}.
Fitting the simulated transmitted intensity decay yields a lifetime of \SI{11.234}{\pico\second}, and we finally obtain $\mu_\text{a} = \SI{0.1997}{\per\milli\meter}$ and $l_\text{t} = \SI{897}{\micro\meter}$, to be compared with $\mu_\text{a} = \SI{0.2}{\per\milli\meter}$ ($\delta x \approx \num{-1.5e-3}$) and $l_\text{t} = l_\text{s}/(1-g) = \SI{45}{\micro\meter}/\num{0.05} = \SI{900}{\micro\meter}$ ($\delta x \approx \num{-3e-3}$).

The second implementation of our routine allows to retrieve $l_\text{t}$ and $g$ assuming that $\mu_\text{a}$ is known. A common case is that of vanishing absorption, which is often encountered when studying for example metal oxide powders with NIR radiation. We therefore reconsider the same iso-level curve of \fref{fig:MCLUT}a which yields the same $\text{OT}^{-1}$. Superimposing this condition over the $\tau(n=\num{1.38})$ surface and its experimental iso-$\tau$ curve of \SI{21.936}{\pico\second} (\fref{fig:isoTau}) finally gives the estimated $g$ parameter, for example by means of spline interpolation over a discrete set of evaluated iso-$D$ $(\text{OT}^{-1}, g)$ pairs (\fref{fig:MCLUT}d). In this case we obtained $l_\text{t} = \SI{897}{\micro\meter}$ and $g = \num{0.938}$ ($\delta x \approx \num{-1.2e-2}$).

The usefulness of our look-up table (LUT) routine is further demonstrated by considering recent experimental data, where mean square width measurements are obtained by means of an ultrafast optical gating technique. Pattelli et al.~\cite{pattelli2015visualizing} recently highlighted the robustness of the mean square width as a valuable experimental observable for the retrieval of the microscopic transport parameters. In that work, the authors consider a homogeneous isotropic sample made of $\text{TiO}_2$ nanoparticles ($g=0.6$, with vanishing absorption at the working wavelength of \SI{810}{\nano\metre}) embedded in a polymer matrix; sample thickness was measured to be \SI{203}{\micro\metre} and the average refractive index at the working wavelength is \num{1.52}, close to that of many biological tissues. The observed experimental lifetime and mean square width slope were $\tau = \SI{6.01}{\pico\second}$ and \SI{6984}{\micro\metre\squared\per\pico\second} and a value of $l_\text{t}=\SI{25.5}{\micro\metre}$ was found with a brute-force Monte Carlo inversion procedure against the experimental data, involving the simulation of many combination of optical parameters. Here, by feeding the same experimental parameters into our look-up table routine, we instantly find a value of $l_\text{t}=\SI{25.7}{\micro\metre}$, in good agreement with the value reported therein.

Thorough evaluation of errors should be performed on a wide range of parameters, both from simulations and experimental data, which is beyond the scope of this work. Nonetheless it is clear that, especially at lower thicknesses where the diffusion approximation is more defective, our routine offers very accurate retrieving capabilities as compared to other slab-geometry fitting and/or LUT approaches \cite{pifferi1998real}. Uncertainties as low as a few percent with respect to simulated data have been demonstrated in other works for the semi-infinite geometry using integrated intensities as the input parameters. It might be questioned whether this kind of uncertainty is realistic, since integrated sphere measurements themselves suffer of both random and systematic errors of similar magnitude in the first place\cite{dam2000multiple}. On the contrary, the slope of the mean square width and the transmittance decay lifetime can be typically determined with better precision, accuracy and robustness, since their scalar value depends on the collective arrangement of multiple points of a curve.

As a last point, it is interesting to discuss on possible extensions of our routine applicability. At least a third input observable in addition to the lifetime and the mean square width slope needs to be known in order to retrieve simultaneously all three transport parameters at once from an unknown medium. A possible candidate could be represented by the asymptotic tail slope of a steady state profile, which should exhibit an appreciable dependence on $g$ at lower optical thicknesses. For all practical purposes, this asymptotic tail slope would feature all the previously listed advantages, with the possible exception of the last one because of the need to add absorption \textit{ex post}.
Other relative parameters could be exploited, taking advantage of their $g$ dependence, such as the transmittance rising time\cite{svensson2013exploiting}, even though such parameters would clearly not benefit from all the aforementioned points.
To sum up, look-up table methods are very general in their nature and consequently can be profitably applied in a number of practical use cases. Of course, in order to tackle more complex samples (e.g.~multilayered or anisotropic slabs) more observables are needed. Nonetheless we believe that, whenever possible, mean square width and lifetime measurements should always be preferred and included in every LUT-based retrieval routine because of their intrinsic robustness.

\section{Methods}
\subsection*{Software}
For the purpose of this work we developed a new multi-layered Monte Carlo C++ software library called MCPlusPlus, whose source code can be freely consulted, used, and contributed at \oursourcecode. This software was developed from scratch aiming at enriching existing multi-layer Monte Carlo software such as MCML\cite{wang1995mcml} or CUDAMCML\cite{alerstam2008parallel}. Being developed entirely in C++, the program takes extensive advantage of object-oriented programming paradigm (OOP), which is particularly suited to model a random walk problem\cite{doronin2011online}. Since pieces of code can be encapsulated in reusable \emph{objects}, OOP offers several advantages including scalability, modularity, ease of maintenance and abstraction. Of equal importance is the fact that OOP naturally lends itself as a tool to describe a high-level \emph{interface} to the software itself. Indeed, MCPlusPlus comes as a shared library rather than an executable package. A Python interface to the library is also provided so that simulations are extremely easy to set up and run through very simple scripts. Scriptability proves indeed to be very useful as it easily allows to loop over the parameter space when building a look-up table or performing a brute-force fit. We believe this to be a key strength of our package which improves considerably on existing multi-layer Monte Carlo software and will hopefully encourage its adoption.

Random walk implementations of light transport fall into the category of so-called ``embarrassingly parallel'' problems, whose solution can largely benefit from the increasing availability of parallel computing architectures such as GPUs (Graphics processing units) and multi-core CPUs. Yet, despite delivering fastest performance, GPUs still present some limitations. 
We therefore developed our software to be run on CPUs in order to better meet our goal. In particular, being our approach based on a look-up table, our main targets are numerical accuracy, reliability and reproducibility rather than execution speed. Indeed, current GPU implementations of the light transport problem in scattering media are more often targeted at realistic biomedical applications involving complex meshes which would otherwise present an overwhelming computational burden, rather than fundamental and statistical studies such those underlying this study.
However we must note that, despite running on CPUs, MCPlusPlus performance is not sacrificed as we can still exploit the ubiquitous multi-core architecture of modern computers; performance close to GPU is soon matched on a small computing cluster or even on a single multi-core workstation. CPU code also ensures maximum hardware compatibility, while GPU-based implementations are hardware or even vendor specific. This should make the adoption of MCPlusPlus as broad as possible so to also encourage further expansion of the dataset on which our look-up table is based. Additionally, developing software for a pure CPU architecture has generally less complications than writing GPU-compatible code; plenty of software libraries and high-quality Pseudo-Random Number Generators (PRNG) are widely available for the CPU, providing us with the flexibility and freedom that we needed for the purpose of this work.

Finally, the magnitude of simulations performed for this and a related study based on the same software\cite{pattelli2015diffusion} is particularly significant, which alone poses several challenges. In particular, performing simulations with a large number of walkers ($\gtrsim \num{e10}$) requires the use of 64-bit PRNGs in place of the more common 32-bit implementations, which would introduce a statistically significant truncation in the step length distribution. Accordingly, the correct representation of 64-bit generated random variates requires the use of long double floating point notation. Both these requirements are straightforward on a CPU architecture, as opposed to GPUs, supporting our preference for the former.

Regarding the simulation output, MCPlusPlus allows to collect statistics on photon exit times, exit points and exit angles. The software provides a powerful histogramming interface that allows to specify any number of simple or bivariate histograms, so that both steady state or time-resolved statistics can be extracted very easily. Nevertheless, if desired it is also possible to obtain raw, non-histogrammed data on a photon-by-photon basis. This can be useful for some particular applications, for example it allows one to introduce photon absorption \emph{after} a single simulation has been run. All output is provided in the widespread HDF5 binary file.

\subsection*{Simulations}
The look-up table (LUT) dataset which we described is based directly on the hyper-surfaces shown in figures \ref{fig:Dcube} and \ref{fig:lifetimecube}, which result from a fine sampling of the ($n$, $g$, $\text{OT}^{-1}$) parameter space for the simple case of a single slab. In particular, all combinations of the following values have been simulated: $n \in [\numlist[list-separator = {;\allowbreak}, list-final-separator = {;\allowbreak}]{0.6; 0.8; 0.9; 0.95; 1; 1.02; 1.05; 1.1; 1.2; 1.3; 1.4; 1.5; 1.6; 1.8; 2.0; 2.2}]$, $g \in [\numlist[list-separator = {;\allowbreak}, list-final-separator = {;\allowbreak}]{0; 0.1; 0.2; 0.3; 0.4; 0.5; 0.6; 0.7; 0.8; 0.9; 0.99}]$, $\text{OT}^{-1}\in [\numlist[quotient-mode = fraction, list-separator = {;\allowbreak}, list-final-separator = {;\allowbreak}]{0.1; 1/9; 1/8; 1/7; 1/6; 0.2; 0.25; 0.3; 0.35; 0.4; 0.5; 0.6; 0.7; 0.8; 0.9; 1}]$. The slab thickness and internal refractive index were kept constant to $L_0=\SI{1000}{\micro\meter}$ and $n_\text{in}=1$ respectively, while varying $l_\text{s}$ and $g$. For each configuration we simulated \num{e9} photons emitted from a pencil-beam source impinging normally on the slab. Photons are propagated inside the scattering material through a standard random-walk algorithm, where scattering lengths follow an exponential distribution and scattering angles are generated using the well-known Henyey-Greenstein function. To build the look-up table, we first fit a single exponential decay to the time-resolved transmitted intensity curve to obtain a first estimate of the lifetime $\tau^\prime$. We then repeat the same fit, this time on the range $4\tau^\prime$--$9\tau^\prime$, to obtain the final estimate for $\tau$. This ensures that the fitting is done at times long enough to extract the asymptotic value of $\tau$ and adds consistency to the fitting method between different simulations.
Once $\tau$ is obtained, we find $D$ by performing a linear fit on the mean square width $w^2(t)$ as a function of time (see \fref{eq:MSW}), limiting the fit to the same $4\tau$--$9\tau$ range; the lower limit is chosen so to always exclude the early-time photons before the onset of the diffusive regime, while the upper limit is to avoid the noise found at very long times due to insufficient statistics. 
It might be called into question whether it is appropriate to use the decaying lifetime as a time unit for the mean square width evolution, since the former is mainly determined by transport properties along the thickness direction, while the latter occurs along the plane. A lifetime-based time range provides indeed a convenient way of defining a consistent, self-tuning fitting window across the whole dataset. This simple choice is also advocated under practical reasons, since the lifetime is undoubtedly the actual temporal unit that eventually dictates --- both in real and numerical experiments --- the signal-to-noise ratio. In this respect, every diffusion coefficient within our simulated phase space has been determined under equal noise conditions.
No less important, limiting our investigation to a long-time window is also relevant under a more technical point of view: i.e.\ it renders irrelevant for all practical purposes the specific choice of both the spatial source distribution and the phase function.

Values obtained for $\tau$ and $D$ for each simulation are eventually arranged in the form of a hyper-surface as shown in figures \ref{fig:Dcube}b and \ref{fig:lifetimecube}b. In order to neutralize the noise originating from statistic fluctuations and fit uncertainty, we consider each simulated $n$-slice separately and smooth the data through a Loess fitting routine (range parameter set to 0.25) as shown in figures \ref{fig:Dcube}c and \ref{fig:lifetimecube}c. Smoothed slices are eventually put back together to perform a cubic interpolation along the refractive index contrast axis to obtain a hyper-surface for $D$ and $\tau$ that can be evaluated for any triplet in the ($n$, $g$, $\text{OT}^{-1}$) parameter space (figures \ref{fig:Dcube}d and \ref{fig:lifetimecube}d). Interpolation has been performed separately on the $n\leq1$ and $n\geq1$ regions of the parameter space due to the sharp first-derivative discontinuity occurring in $n=1$.

\section{Conclusions}
In this paper we studied the radiative transfer problem in the infinitely extended slab geometry and compared its exact numeric solution (obtained by means of Monte Carlo simulation) to the predictions obtained by diffusion theory. For the first time to our knowledge, both transverse and axial transport are addressed using respectively the mean square width growth rate and the transmittance decay lifetime as figures of merit. These two observables are to be measured at late times, thus providing valuable insight on light transport properties well into the diffusive regime.
Our investigation provides a complete characterization of how the diffusive approximation gradually fails over a range of optical thicknesses from 10 to 1 --- a range where resorting to the diffusion approximation is a questionable yet common practice --- considering both the refractive index contrast and the scattering anisotropy. The mean square width growth rate is of particular interest since its standard prediction according to the diffusion approximation depends solely on the diffusion coefficient, as opposed to other time-resolved observables where also the thickness and refractive index contrast usually play a critical role.

As regards the mean square width expansion rate, we found that deviations from the simple diffusion approximation prediction exist especially at low optical thicknesses and scattering anisotropy, and that they are always in the form of an under-estimation of the actual rate. Notably, when considering the case of high scattering anisotropy which is most relevant in many applications, the magnitude of the observed deviation remains limited even at low optical thicknesses as compared to the lifetimes, which instead deviate significantly even including extrapolated boundary conditions.

In cases where quantitative accuracy is critical or the sample is optically thin, non-approximated methods should be adopted. In this context, readily available look-up tables (LUT) offer a convenient strategy to obviate the computational cost of Monte Carlo simulations. Based on the large simulated dataset, we presented a novel look-up table based on the combination of the decay lifetime and mean square width slope as input parameters, which offer a series of relevant advantages over existing LUT solutions. Prominently, they can be measured without the need of any absolute intensity measurement nor time axis origin determination. On the simulation side, the look-up table is easily built since both the lifetime and the mean square width slope are unaffected by binning strategies, plus there is no need to simulate absorption nor add it subsequently in contrast with previously available solutions.
The creation of a custom look-up table is further streamlined by taking advantage of a fully scriptable Monte Carlo software, which we developed. Both a documented version of the software and the presented look-up table are freely available at \oursourcecode\ and \ourhomepage\ respectively, which we hope will encourage feedback, adoption and contribution.

\ack
This work is financially supported by the European Network of Excellence Nanophotonics for Energy Efficiency and the ERC through the Advanced Grant PhotBots, project reference 291349 funded under FP7-IDEAS-ERC. CT and GM acknowledge support from the MIUR program Atom-based Nanotechnology and from the Ente Cassa di Risparmio di Firenze with the project GRANCASSA.

\section*{References}
\bibliographystyle{unsrt}
\bibliography{references}
\end{document}